\documentclass[12pt]{amsart}

\usepackage[utf8]{inputenc}
\usepackage[T1]{fontenc}
\usepackage{lmodern}

\usepackage{longtable}
\usepackage{multicol}
\usepackage{bm}
\usepackage{fancyvrb}
\usepackage[colorlinks=true,citecolor=blue,hypertexnames=false]{hyperref}
\usepackage{color}

\theoremstyle{plain}
\newtheorem{lem}{Lemma}
\newtheorem*{qst}{Question}

\newcommand{\Diag}{\mathrm{Diag}}
\newcommand{\disc}{\mathrm{disc}}

\newcommand{\bC}{\mathbb{C}}
\newcommand{\bK}{\mathbb{K}}
\newcommand{\bN}{\mathbb{N}}
\newcommand{\bQ}{\mathbb{Q}}
\newcommand{\bR}{\mathbb{R}}
\newcommand{\bZ}{\mathbb{Z}}

\newcommand{\cM}{\mathcal{M}}

\newcommand{\ds}{\partial_s}
\newcommand{\dt}{\partial_t}
\newcommand{\du}{\partial_u}
\newcommand{\dv}{\partial_v}
\newcommand{\dx}{\partial_x}
\newcommand{\duone}{\partial_{u_1}}
\newcommand{\dutwo}{\partial_{u_2}}

\def\pFqnoargs#1#2{{}_#1F_#2}
\def\pFq#1#2#3#4#5#6{\pFqnoargs{#1}{#2}\biggl(\begin{matrix}%
{#3}\kern.707em{#4}\\{#5}%
\end{matrix}\,\bigg|\,#6\biggr)}

\title[Explicit formula for diagonal {3D} rook paths]{Explicit formula for the generating series of diagonal {3D} rook paths}
\author[A. Bostan]{Alin Bostan$\mbox{}^\dagger$}\thanks{$\dagger$ Supported in part by the Microsoft Research\,--\,INRIA Joint Centre}
\address{INRIA (France)}
\email{alin.bostan@inria.fr}
\author[F. Chyzak]{Frédéric Chyzak$\mbox{}^\dagger$}
\address{INRIA (France)}
\email{frederic.chyzak@inria.fr}
\author[M. van Hoeij]{Mark van Hoeij$\mbox{}^\ddag$}\thanks{$\ddag$ Supported by NSF 1017880}
\address{Florida State University, Department of Mathematics
(Tallahassee, FL, USA)}
\email{hoeij@math.fsu.edu}
\author[L. Pech]{Lucien Pech$\mbox{}^\dagger$}
\address{École Normale Supérieure (France)}
\email{lucien.pech@gmail.com}

\date{}
\begin{document}

\begin{abstract}
Let $a_n$ denote the number of ways in which a chess rook can move from a
corner cell to the opposite corner cell of an $n \times n \times n$
three-dimensional chessboard, assuming that the piece moves closer to the goal
cell at each step. We describe the computer-driven \emph{discovery and
proof\/} of the fact that the generating series $G(x)= \sum_{n \geq 0}
a_n x^n$ admits the following explicit expression in terms of a Gaussian hypergeometric function:
\[
G(x) =  1 + 6 \cdot \int_0^x \frac{ \,\pFq21{1/3}{2/3}{2}
  {\frac{27 w(2-3w)}{(1-4w)^3}}}{(1-4w)(1-64w)} \, dw.
\]
\end{abstract}

\maketitle

\section{Introduction}

In this article, we solve a problem of enumerative combinatorics addressed and
left open in~\cite{ErFeTr10}. The initial question is formulated in terms of
paths on an infinite three-dimensional chessboard. The 3D chessboard being
identified with $\bN^3$, a 3D rook is a piece which is allowed to move
parallelly to one of the three axes. The general objective is to count paths
(i.e., finite sequences of moves) of a 3D rook on the 3D chessboard.
Following~\cite{ErFeTr10}, we further restrict to 3D rook paths starting from
the cell $(0,0,0)$, whose steps are positive integer multiples of
$(1,0,0)$, $(0,1,0)$, or $(0,0,1)$. In other words, we assume that the piece
moves closer to the goal cell at each step. For example, one such rook path is
\[(0,0,0) \rightarrow (5,0,0) \rightarrow (5,0,6) \rightarrow (10,0,6)
\rightarrow (10,2,6). \]

Let $r_{i,j,k}$ denote the number of rook paths from $(0,0,0)$ to $(i,j,k)\in \bN^3$,
and let $a_n=r_{n,n,n}$ be the number of diagonal rook paths from $(0,0,0)$ to
$(n,n,n)$. The sequence $(a_n)$ is called the 3D \emph{diagonal rook
sequence\/}; its first terms are
\begin{multline*}
1, 6, 222, 9918, 486924, 25267236, 1359631776,\\
75059524392, 4223303759148,\ldots
\end{multline*}
This is Sequence A144045 in
Sloane's on-line encyclopedia of integer sequences at
\url{http://oeis.org/A144045}. It was conjectured in~\cite{ErFeTr10} that the sequence $(a_n)$ satisfies the
fourth-order recurrence relation
\begin{multline}\label{eq:recErickson}
2n^2(n-1)a_n -(n-1)(121n^2-91n-6)a_{n-1} \\
- (n-2) (475 n^2-2512n+2829)a_{n-2}
  + 18(n-3)(97n^2-519n+702)a_{n-3} \\
- 1152 (n-3) (n-4)^2a_{n-4}= 0,  \qquad \text{for~$n\geq 4$.}
\end{multline}

The aim of the present article is not only to give a computer-driven proof of this conjecture, but also to show that the generating function $G(x) = \sum_{n \geq 0} a_n x^n \in \bQ[[x]]$ admits the following explicit expression:
\begin{equation}\label{eq:G-as-2F1}
G(x) =  1 + 6 \cdot \int_0^x \frac{ \,\pFq21{1/3}{2/3}{2}{\frac{27 w(2-3w)}{(1-4w)^3}}}{(1-4w)(1-64w)} \, dw,
\end{equation}
where
the Gaussian hypergeometric series~$\pFqnoargs21$ with parameters $a,b,c \in \bC$, \ $-c \notin \bN$,
is defined by
\[\pFq21{a}{b}{c}{z} = \sum_{n=0}^\infty \frac{(a)_n(b)_n}{(c)_n} \, \frac
{z^n} {n!},\]
using the notation $(a)_n$ for the Pochhammer symbol $(a)_n=a(a+1)\cdots(a+n-1)$.
%%%%%%%%%%%%%%%%%%%%%%%%%%%%%%%%%%%%%%%%%%%%%%%%%%%%%%%%%%%%%%%%%%%%%%%%%%%%

\bigskip

The core of this article is Section~\ref{sec:solution}, in which we
complete the enumerative study of 3D~rooks, proving in particular
recurrence~\eqref{eq:recErickson} and the explicit
form~\eqref{eq:G-as-2F1}.  Section~\ref{sec:further} then presents
further comments, variant proofs and calculations, and discussions of
a possible extension.

In more detail, Section~\ref{sec:solution} proceeds as follows.
The question of the enumeration of a constrained combinatorial walk is
re-expressed in Section~\ref{sec:from-comb-to-alg} in terms of a
diagonal of a rational series.  Then, this diagonal is encoded as a
residue in Section~\ref{sec:key-eq}, leading to the problem of finding
a special ODE it satisfies.  In the literature, an effective approach
to finding this special ODE was developed by Lipshitz.  In
Section~\ref{sec:lipshitz}, we show that, while the approach applies
in theory, it leads to unfeasible calculations in practice.  A faster
approach, Zeilberger's creative telescoping and its evolution by
Chyzak, is summarised and applied in Section~\ref{sec:ct-approach}.
The obtained ODE allows us to prove the conjectured
recurrence~\eqref{eq:recErickson} and even a lower-order one,
\eqref{eq:shortrec}, in Section~\ref{sec:proving-rec}.  We then derive
the $\pFqnoargs21$~explicit form in
Section~\ref{sec:explicit-form-sol}.

When possible, we have attempted to describe the mathematical
phenomena, ideas, and techniques on a general level in
Sections~\ref{sec:from-comb-to-alg}--\ref{sec:explicit-form-sol}.
When the complete proofs of the results we state require a more
involved calculation, and also when they are 
based on more involved notions or on the fine tuning of
algorithms, we have postponed some of the technicalities to
Appendix~\ref{sec:ct-appendix}.

\section{Main steps of the solution}\label{sec:solution}

\subsection{From combinatorics to algebra}\label{sec:from-comb-to-alg}
The combinatorial problem translates into a question on generating series, best expressed in terms of \emph{diagonals\/}:
given $\phi(s,t,u) = \sum_{i,j,k \geq 0} c_{i,j,k} s^i t^j u^k$
in~$\bQ[[s,t,u]]$, its \emph{diagonal\/} is
defined as the univariate power series $\Diag(\phi) = \sum_{n\geq 0} c_{n,n,n}
x^n$ in $\bQ[[x]]$.

\begin{qst}
Define the trivariate rational function $f(s,t,u)$ in
$\bQ(s,t,u)$ as
\begin{multline*}
\left( 1 - \sum_{n \geq 1} s^n - \sum_{n \geq 1} t^n  - \sum_{n \geq 1} u^n
\right)^{-1} = \\
\frac{(1-s)(1-t)(1-u)}{1-2(s+t+u)+3(st+tu+us)-4stu} ,
\end{multline*}
which is seen to be at the same time in $\bQ[[s,t,u]]$.
What can be said about its diagonal $\Diag(f) = \sum_{n\geq 0} a_{n} x^n$?
\end{qst}

This translation is based on the following straightforward extension
of Proposition~6.3.7 in~\cite{Stanley99},
which we shall use with the choice
$\mathfrak{S} = \{\, (n, 0, 0) :
n \geq 1 \,\} \cup \{\, (0, n, 0) : n \geq 1 \,\} \cup \{\, (0, 0, n) : n \geq 1 \,\}.$

\begin{lem}\label{basiclemma}
Given a finite set of allowed steps $\mathfrak{S} \subseteq \bN^3 \setminus \{(0,0,0)\}$, let
$N_\mathfrak{S}(m,n,p)$ denote the number of paths from $(0,0,0)$ to $(m,n,p)$, with path steps taken from~$\mathfrak{S}$.
Then, the generating series
\[ G_\mathfrak{S}(s,t,u) = \sum_{m,n,p \geq 0 } N_\mathfrak{S}(m,n,p) s^m t^n u^p \]
is rational, and has the form
\[ G_\mathfrak{S}(s,t,u) = \frac{1}{1 - \sum_{(i,j,k) \in \mathfrak{S}} s^i t^j u^k} . \]
The same result holds for infinite step sets~$\mathfrak{S}$ provided the
power series $\sum_{(i,j,k) \in \mathfrak{S}} s^i t^j u^k$ is rational.
\end{lem}

\subsection{The key equation}\label{sec:key-eq}

A first, theoretical answer to our question comes from the main result in
Lipshitz's article~\cite{Lipshitz88}: the diagonal $\Diag(f)$ is a D-finite
series, i.e., it satisfies a linear differential equation in $\dx =
d/dx$ with polynomial coefficients in~$x$. Therefore, the sequence $(a_n)$ of
its coefficients satisfies a linear recurrence with polynomial
coefficients in~$n$.

%[this result generalises the classical result saying that the diagonal of a bivariate rational function is algebraic].

Moreover, let $F$ be the trivariate rational function
\begin{multline*}
F = \frac{1}{st} \cdot f(s, t/s, x/t) = \\
\frac{(s-t)(s-1)(t-x)}{st(-st+2s^2t+2t^2+2xs-3st^2-3xt-3xs^2+4xst)}.
\end{multline*}
Then, Remark~3 in~\cite{Lipshitz88} shows that

\begin{lem}\label{lem:Lipshitz}
If the equation
\begin{equation}\label{eq:trivariate-rat-ct}
P(F) = \frac{\partial S}{\partial s} + \frac{\partial T}{\partial t}
\end{equation}
admits a solution $(P,S,T)$ where $P(x,\dx)$ is a non-zero linear differential
operator with polynomial coefficients in $\bQ[x]$, and $S$ and $T$ are two
rational functions in $\bQ(x,s,t)$, then $P(x,\dx)$ annihilates $\Diag(f)$.
\end{lem}

Lemma~\ref{lem:Lipshitz} is the heart of the result in~\cite{Lipshitz88}, and
the basis of all methods for effectively computing differential equations
satisfied by diagonals.
For completeness, we give here a self-contained proof, which closely
follows~\cite{Lipshitz88}.

\begin{proof} Although $F$ is not a formal (Laurent) power series in $x,s,t$,
it does come from a formal power series after a change of variables, and thus
it can be seen as an element of the differential $\bQ[x,s,t]$-module of
all formal sums of the form
\[B = \sum_{{\alpha\in
\bN,\beta\in\bZ,\gamma\in\bZ} \atop {\alpha+\beta \geq
-\mu, \beta+\gamma \geq -\nu}} c_{\alpha,\beta,\gamma} s^\alpha t^\beta
x^\gamma\]
for some $(\mu,\nu) \in \bN^2$ depending on~$B$. As a
consequence, it makes sense to speak of the series expansion of $F$ and of
$P(F)$, but in later calculations, these series may be multiplied solely by elements of $\bQ[x,s,t]$.

The diagonal $\Diag(f)$ is, by construction, equal to the coefficient
of~$s^{-1} t^{-1}$ in~$F$.  On the other hand,
equation~\eqref{eq:trivariate-rat-ct} implies that the coefficient
of~$s^{-1} t^{-1}$ in~$P(F)$ is zero.  The conclusion then follows
from the sequence of equalities
\[
P(\Diag(f))
= P([s^{-1} t^{-1}] F)
=  [s^{-1} t^{-1}] P(F)
= 0, \]
the second equality being a consequence of the fact that $P$ is free of $\ds$ and $\dt$.
\end{proof}

The question remaining is how to determine a non-trivial triple $(P,S,T)$ solution of~\eqref{eq:trivariate-rat-ct}.

\subsection{Lipshitz's approach}\label{sec:lipshitz}

Lipshitz~\cite[Lemma~3]{Lipshitz88} uses a counting argument,
similar to Fasenmyer's~\cite{Fasenmyer47,Zeilberger82}, to show
that there exists a non-zero linear differential operator $L(x,\dx,\ds,\dt)$,
free of $s$ and~$t$, that annihilates $F$.
Then, writing this operator under the form
\begin{multline*}
L(x,\dx,\ds,\dt) = P(x,\dx) + \\
(\text{higher-order terms in $\ds$ and~$\dt$, free of $s$ and~$t$}),
\end{multline*}
it follows from Lemma~\ref{lem:Lipshitz} that $P$~annihilates~$\Diag(f)$.

Let us sketch Lipshitz's argument on our concrete example;
as we shall see,
while this proof is algorithmic and ultimately reduces the existence of $P$ to
a linear-algebra argument, it produces unreasonably large systems. The key
observation is that, by Leibniz's rule, for any nonnegative integer $N$, the
$\binom{N + 4}4$ rational functions
\[ x^i \dx^j \ds^k \dt^\ell (F),  \quad 0 \le i+j+k+\ell \leq N, \]
belong to the $\bQ$-vector space of dimension at most
$18 (N+1)^3$ spanned by the elements
\begin{equation}\label{eq:mon-over-pow-of-q}
\frac{x^i s^j t^k}{q^{N+1}},\quad  0\le i \le 2N+ 1,\ 0\le j \le 3N+2,
\ 0\le k \le 3N+2,
\end{equation}
where $q$ is the denominator $st(-st+2s^2t+2t^2+2xs-3st^2-3xt-3xs^2+4xst)$
of~$F$. Thus, if $N$ is an integer such that $\binom{N + 4}4 > 18 (N+1)^3$,
then there exists a non-zero linear differential operator $L(x,\dx,\ds,\dt)$
of total degree at most $N$ in $x$, $\dx$, $\ds$, and $\dt$, such that $L (F)
= 0$.

Now, it appears that $N=425$ is the smallest integer satisfying the inequality
$\binom{N + 4}4 > 18 (N+1)^3$; therefore, explicitly finding the operator
$L$ would require solving a homogeneous linear system with some 1,391,641,251 unknowns and
some 1,391,557,968 equations!

It is possible to refine the previous argument in various ways, and thus
to reduce the 
search of $L$ to a linear algebra problem of
smaller size.
Such methods will be discussed
and compared
thoroughly in a later work.
In our case, the best reduction in size that we could
achieve uses a variation of the main idea in~\cite[\S3.2]{BoChLeSaSc07}.

First, as before, Leibniz's rule shows that the $u_N := \binom{N + 3}3$ rational
functions
\[ \dx^j \ds^k \dt^\ell (F), \quad 0 \le j+k+\ell \leq N, \]
belong to the $\bQ(x)$-vector space spanned by the elements
\begin{equation}\label{eq:biv-over-pow-of-q}
\frac{s^j t^k}{q^{N+1}},\quad  0\le j \le 3N+2,
\ 0\le k \le 3N+2,
\end{equation}
whose dimension~$d_N$ is at most~$9 (N+1)^2$.

A further idea is to exploit the sparse feature of the
monomial supports of the
numerator and denominator of $F$. By studying the behaviour of
these supports under
differentiations with respect to $x$, $s$, and $t$, one shows that
the
terms~\eqref{eq:biv-over-pow-of-q} will not be involved for
the $\binom{2N+2}2 + \binom{N+2}2$ 
pairs
~$(j,k) \in \mathbb{N}^2$ in the set
\begin{multline*}
\{\, (j,k) : j+k \leq 2N \,\}
\quad \cup \quad \\
\{\, (j,k) : 2N+2\leq j, k\leq 3N+2, \ j+k \geq 5N+4 \,\}.
\end{multline*}

This implies that $d_N$~is bounded above by $e_N := \frac12 (13N+14)(N+1)$,
and shows that the search of $L$ boils down to a homogeneous linear system
over $\mathbb{Q}(x)$, with $u_N$ unknowns and $e_N$ equations. The smallest
value of $N$ such that $u_N > e_N$ is $N=36$, leading to the task of computing
the (right) kernel of a matrix of size $8917\times 9139$, with entries in
$\mathbb{Q}[x]$ of degree at most 37.

Despite the spectacular reduction in dimension compared to Lipshitz's original
argument, these sizes are still too high to be dealt with in practice. Of
course, the value $e_N$ is only an upper bound on $d_N$, leaving room
for a magical rank drop occurring before the predicted value $N=36$.
That being said, our implementation of the method constructs the linear system
for increasing values of~$N$, but does not find any non-trivial kernel for
$N\leq 18$, and is not able to finish the computation for $N=19$
before consuming all the available memory.\footnote{Our implementation of an optimisation of Lipshitz's algorithm, available at \url{http://algo.inria.fr/chyzak/Rooks/OptimisedLipshitz.mpl}, is nevertheless able to predict that, with high probability, a non-trivial kernel does exist for $N=19$.}

The conclusion is that Lipshitz's approach and its improvements are not
sufficient to obtain effectively an equation for the diagonal of~$f(x,s,t)$.

%%%%%%%%%%%%%%%%%%%%%%%%%%%%%%%%%%%%%%%%%%%%%%%%%%%%%%%%%%%%%%%%%%%%%%

\subsection{The creative-telescoping approach}\label{sec:ct-approach}

An alternative approach to solving~\eqref{eq:trivariate-rat-ct}
was popularised among the computer-algebra community
by Zeilberger under the name of \emph{creative telescoping}, and it is
the one that we shall use to effectively construct an equation for our
diagonal.
This method, first introduced for hypergeometric
summation~\cite{Zeilberger:1990:FAP,Zeilberger:1991:MCT} and
hyperexponential integration~\cite{Almkvist:1990:MDI}, was later
generalised by Chyzak~\cite{Chyzak:2000:EZF} to more general integrands involving a larger
class of special functions called
\emph{holonomic}.
Zeilberger's definition of this
class~\cite[Section~2.2.4]{Zeilberger-1990-HSA} has its roots in
D-module theory~\cite{Bernstein71,Bernstein72} and is not needed to
understand the present paper.
See also~\cite{Chyzak:2009:NHS} for recent extensions of creative telescoping to more general
special functions.

Prior to summarising the method, we show its behaviour and its result
on our concrete case, postponing the explicit details of intermediate
steps to Appendix~\ref{sec:ct-appendix}.

\subsubsection{Obtaining the result}\label{sec:the-result}

In our case, Chyzak's algorithm, as implemented by the package \textsf{Mgfun}\footnote{%
version 4.0, available as part of Algolib~13.0 from
\url{http://algo.inria.fr/libraries/}.
Initially developed by F.~Chyzak.
Specific version used here reimplemented by L.~Pech.}
for \textsf{Maple}, produces a solution $(P,S,T)$ of~\eqref{eq:trivariate-rat-ct},
with $P = P_2 \dx$, where
\begin{multline}\label{eq:P2}
P_2 = x(x-1)(64x-1)(3x-2)(6x+1)\dx^2 \\
+(4608x^4-6372x^3+813x^2+514x-4)\dx \\
+4(576x^3-801x^2-108x+74).
\end{multline}

Let $q = st \cdot q_1$ with $q_1$ irreducible in $\bQ[x,s,t]$ of degree
$(1,2,2)$ in $(x,s,t)$. Then, $S$ and $T$ are of the very special forms:
\[S = \frac{(s-t) \cdot U}{2st \cdot q_1^2 \cdot \disc_t(q_1)}, \quad T =
\frac{(s-t) \cdot V}{2s^2 \cdot q_1^3 \cdot \disc_t(q_1)^2},\]
where $U$ and~$V$ are (irreducible) polynomials in $\bQ[x,s,t]$ of
respective degrees $(5,8,3)$ and $(8,14,5)$ in $(x,s,t)$. Here,
$\disc_t(q_1)$ denotes the discriminant of $q_1$ with respect to $t$.
Observe the very moderate degrees in comparison to those
pessimistically predicted by Lipschitz's method.

\textsf{Maple} commands to produce $P$, $S$, and~$T$ are provided online
at \url{http://algo.inria.fr/chyzak/Rooks/CreativeTelescoping.mpl}; \,
they take less than 10~seconds on an average laptop.
The rational functions $S$ and~$T$ are not displayed here, but can be
found online at
\url{http://algo.inria.fr/chyzak/Rooks/Certificates.mpl}.

\subsubsection{Sketch of the algorithm}\label{sec:sketch}

We proceed to sketch the algorithm of~\cite{Chyzak:2000:EZF} for the
creative telescoping of multiple “integrals.”  By this, we actually
mean in this article answering the algebraic question of how to solve the key
equation, whether in the form~\eqref{eq:bivariate-rat-ct} below for
“simple integrals,” or in the form~\eqref{eq:trivariate-rat-ct} for
“double integrals.”

Presenting Chyzak's algorithm directly in its generality would prove
too arid, so we go instead progressively from the simpler case of
rational simple integrals to the integration of more general
functions, then to double integrals.  In effect, this section is
mostly bibliographic, with the exception that, in the case of multiple
integrals, we provide an extension of the algorithm
in~\cite{Chyzak:2000:EZF} beyond the so-called “natural boundaries.”

For a single integral of a bivariate rational function~$F(u,v)$, the
creative-telescop\-ing method finds a linear differential
operator~$P(u,\du)$ and a rational function~$S$ in~$\bQ(u,v)$ such
that (cf.~equation~\eqref{eq:trivariate-rat-ct})
\begin{equation}\label{eq:bivariate-rat-ct}
P(F) = \frac{\partial S}{\partial v}.
\end{equation}
To this end, the algorithm in~\cite{Almkvist:1990:MDI} writes $P$
and~$S$ in undetermined form
\begin{equation}\label{eq:P-Ansatz-1}
P = \eta_r(u) \du^r + \dots + \eta_0(u)
\qquad\text{and}\qquad
S = \phi(u,v) F
\end{equation}
for some tentative order~$r$ and unknown rational functions~$\eta_i$
from~$\bQ(u)$ and $\phi$~from~$\bQ(u,v)$.
To get~$\phi$, the original algorithm of~\cite{Almkvist:1990:MDI}
exploits the specific nature of~$F$ (hyperexponential).
We prefer describing here the less optimal variant
of~\cite{Chyzak:2000:EZF}, as it will extend to the more general class
of inputs we need:
\eqref{eq:bivariate-rat-ct}~is rewritten into an auxiliary
\emph{ordinary\/} linear differential equation on~$\phi$, before one
uses known algorithms for solving a linear differential equation with
polynomial coefficients for its \emph{rational\/} solutions.
More precisely, such algorithms, like the one
in~\cite{Abramov:1991:FAS}, have to be refined to solve for
the~$\eta_i$ as well;
this is easily taken into account just by having more unknowns in the
crucial step of linear algebra over~$\bQ(u)$ in this rational solving.

For reasons to be given shortly, the double integration of trivariate
\emph{rational\/} functions is based on integration of more general
functions~$F$, leading to a more general form for~$S$
in~\eqref{eq:bivariate-rat-ct}.
Specifically, the case dealt with in~\cite{Chyzak:2000:EZF} is that of
functions~$F$ for which there exists a finite maximal set of
derivatives~$\du^a\dv^b(F)$ that are linearly independent
over~$\bQ(u,v)$.
Such functions are known as \emph{differentiably finite\/} in the
literature, or in short \emph{D-finite\/}
\cite{Stanley-1980-DFP,Lipshitz-1989-DFP}.
However, note that the notions of holonomic and differentiably finite
functions, which look technically very different, cover in fact
exactly the same sets of functions.  This fact is explained in
elementary terms in~\cite[Appendix]{Takayama92}.  Still, the more
elaborate holonomic point of view is needed in proofs that require a
subtler counting argument.

Once such a set of pairs~$(a,b)$ is fixed for a given D-finite function~$F$,
equation~\eqref{eq:bivariate-rat-ct}~is changed by setting~$S$ to the
undetermined form
\begin{equation}\label{eq:S-as-finite-sum}
S = \sum_{(a,b)} \phi_{a,b}(u,v) \du^a\dv^b(F)
\end{equation}
for unknown rational functions~$\phi_{a,b}$ from~$\bQ(u,v)$.
The only change then is that the auxiliary equation in~$\phi$ becomes an
auxiliary ordinary linear differential \emph{system\/} in the~$\phi_{a,b}$'s.
Solving is performed either by uncoupling and using the algorithm
mentioned previously several times, or by a direct approach as in~\cite{Barkatou99}.

The case of double integrals leads to
generalising~\eqref{eq:bivariate-rat-ct}
into~\eqref{eq:trivariate-rat-ct}, but the solving does not generalise
so smoothly:
attempting to write $S=\phi_1(s,t,x)F$ and~$T=\phi_2(s,t,x)F$ for
undetermined rational functions $\phi_1$ and~$\phi_2$
from~$\bQ(s,t,x)$ yields a linear \emph{partial\/} differential
equation relating these functions with $\partial\phi_1/\partial s$
and~$\partial\phi_2/\partial t$.
To the best of our knowledge, although this overdetermined equation
has a very specific form, no algorithm is available to solve it for
its rational solutions.

Therefore, instead of a direct approach, Chyzak developed
in~\cite{Chyzak:2000:EZF} a cascading approach which we now summarise.
Noting that the dependency of~$P$ on a \emph{single\/}
derivation~$\du$ in~\eqref{eq:bivariate-rat-ct} is inessential, the
same approach is possible for the creative telescoping with respect to
the (single) variable~$v$ of a \emph{trivariate\/} rational
function~$F$ from~$\bQ(u_1,u_2,v)$.
Indeed, setting~$P$ to the undetermined form
\begin{equation}\label{eq:P-Ansatz-2}
P = \sum_{0 \leq i+j \leq r} \eta_{i,j}(u_1,u_2) \duone^i\dutwo^j
\end{equation}
for some tentative total order~$r$ and unknown rational
functions~$\eta_{i,j}$ from~$\bQ(u_1,u_2)$ and performing the same
solving as previously, now relying on linear algebra over
$\bQ(u_1,u_2)$, leads to a basis of~$P^{(\alpha)}$'s of total order at
most~$r$ for which there exists a rational function
$\phi^{(\alpha)}(u_1,u_2,v)$ satisfying
\[ P^{(\alpha)}(F) = \frac{\partial}{\partial v}\left(\phi^{(\alpha)}F\right). \]
The theory (as developed, e.g., by Zeilberger in~\cite{Zeilberger-1990-HSA})
guarantees that the function~$F$ is D-finite and that
the set of~$P^{(\alpha)}$'s obtained for sufficiently large~$r$
can be used to determine the \emph{finite\/} set
needed as an input to the algorithm of~\cite{Chyzak:2000:EZF} and
in~\eqref{eq:S-as-finite-sum}.

Finally, a double integration algorithm is obtained by continuing the
approach used for natural boundaries in~\cite{Chyzak:2000:EZF} (Stages
A and~B below) by a suitable recombination of the outputs (Stage~C
below).  The resulting treatment of multiple integrals over
non-natural boundaries is an extension over~\cite{Chyzak:2000:EZF},
and the corresponding algorithm is as follows:
\begin{itemize}
\item\emph{Stage~A: First iteration of creative telescoping.}
Using the univariate algorithm for trivariate rational functions
and variables $(s,x,t)$ in place of $(u_1,u_2,v)$ delivers identities
\begin{equation}\label{eq:identities-after-first-iteration}
P^{(\alpha)}(s,x,\ds,\dx)(F) =
  \frac{\partial}{\partial t}\left(\phi^{(\alpha)}(s,t,x)F\right) .
\end{equation}
\item\emph{Stage~B: Second iteration of creative telescoping.}
Considering a function~$\hat F$ of $(s,x)$ that is
annihilated by all~$P^{(\alpha)}$ and using the univariate algorithm
for general functions and variables~$(s,x)$ in place of~$(u,v)$ delivers an identity
\begin{equation}\label{eq:identities-after-second-iteration}
P(x,\dx)(\hat F) =
  \frac{\partial}{\partial s}\bigl(Q(s,x,\ds,\dx)(\hat F)\bigr) .
\end{equation}
\item\emph{Stage~C: Recombination.}
By the theory of linear-differential-operators ideals,%
\footnote{%
This observation was made during a discussion between F.~Chyzak,
M.~Kauers, and Ch.~Koutschan on July 17, 2008, in the ``UFO'' room
at RISC (Linz, Austria).}%
\ the calculations of the algorithm can be interpreted as a proof of
existence of operators~$L^{(\alpha)}(s,x,\ds,\dx)$ satisfying
\begin{equation}\label{eq:identity-after-two-iterations}
P(x,\dx) - \ds Q(s,x,\ds,\dx) = \sum_\alpha
L^{(\alpha)}(s,x,\ds,\dx) P^{(\alpha)}(s,x,\ds,\dx) .
\end{equation}
These~$L^{(\alpha)}$ can effectively be obtained either by following
the calculations step by step or (less efficiently) by a
postprocessing (non-commutative multivariate division).
Hence, defining
\begin{equation}\label{eq:recombining-S-and-T}
S = Q(s,x,\ds,\dx)(F)
\quad\text{and}\quad
T = \sum_\alpha L^{(\alpha)}(s,x,\ds,\dx)
\bigl(\phi^{(\alpha)}(s,t,x)F\bigr)
\end{equation}
leads to a solution~$(P,S,T)$ of~\eqref{eq:trivariate-rat-ct}.
\end{itemize}

Note that this two-stage process inherently introduces a dissymmetry
in the treatment of the variables $s$ and~$t$:
the output from the first iteration tends to be larger than its input;
in turn, the output from the second is larger than the output from the
first.
As a consequence, the order we deal with the variables may have an
impact on the running time.

\subsection{Proving the recurrence}\label{sec:proving-rec}
Translating the differential equation $P(\Diag(f))=0$ into a recurrence
satisfied by the coefficients $a_n$ of $\Diag(f)$ yields the fourth-order
recurrence relation~\eqref{eq:recErickson}.

It is possible to prove that the sequence $(a_n)$ satisfies the
even shorter recurrence
\begin{multline}\label{eq:shortrec}
2(n-1)(35n-52)n^2a_n
-(n-1)(4655n^3-11781n^2+8494n-1776)a_{n-1} \\
+(n-2)(11305n^3-41856n^2+46487n-13128)a_{n-2} \\
-192(n-3)^2(35n-17)(n-2)a_{n-3} =0,
\qquad \text{for~$n \geq 3$.}
\end{multline}

Recurrence~\eqref{eq:shortrec} can be \emph{guessed\/} (e.g., by
linear algebra) using, say, the first 25 terms of the
sequence~$(a_n)$. Once it is conjectured, proving its correctness is
easy:
if $(b_n)$ denotes the sequence defined by the left-hand side
of~\eqref{eq:shortrec}, then recurrence~\eqref{eq:recErickson} implies
that $(b_n)$ satisfies the recurrence
\begin{multline*}
b_n+ 6 b_{n-1} =
  (35n-52) \Bigl(
  2(n-1)n^2a_n
  -(n-1)(121n^2-91n-6)a_{n-1} \\
-(n-2)(475n^2-2512n+2829)a_{n-2}
  +18(n-3)(97n^2-519n+702)a_{n-3} \\
-1152(n-3)(n-4)^2a_{n-4}
\Bigr) = 0,
\qquad \text{for~$n \geq 3$,}
\end{multline*}
which, combined with
the initial condition $b_0 = -54864  \cdot 6  + 2 \cdot 43362 \cdot 222 - 2\cdot 2  \cdot 53  \cdot 3^2\cdot 9918 = 0$, yields $b_n=0$ for all $n \geq 0$.

Maple code to obtain the first terms of the sequence and derive the
recurrences can be obtained at
\url{http://algo.inria.fr/chyzak/Rooks/Recurrence.mpl}.

\subsection{Solution in explicit form}\label{sec:explicit-form-sol}

Operator $P_2$ from equation~\eqref{eq:P2} has a power series solution
with integer coefficients, namely $\dx(\Diag(f))$.
By a speculation of Dwork's~\cite{Dwork90}, it appears plausible that
$P_2$ has a $\pFqnoargs21$-type solution.
We shall briefly sketch how, for a second-order operator~$L$, one
might find such solutions.
A similar problem (for Bessel instead of $\pFqnoargs21$) was solved
in \cite{debeerst, yuan}, and many of the details discussed in those papers are
needed for the $\pFqnoargs21$-case as well (in particular the first
part of Section~3 in \cite{debeerst}).
The result of our computation is:
\begin{equation} \label{eq:closedform}
\dx(\Diag(f)) =
\frac{6}{(1-4x)(1-64x)} \cdot \pFq21{1/3}{2/3}{2} { \frac{27
x(2-3x)}{(1-4x)^3}}.
\end{equation}

\subsubsection{Formal asymptotic-series solutions of linear
  differential operators}

We recall the necessary facts from the classical theory of
asympto\-tic-series solutions of linear differential operators
\cite{Ince:1956:ODE,Wasow:1987:AEO}, and specialise them here to our
need: the order~2.

With respect to a given linear differential operator~$L$ of order~$r$
with polynomial coefficients, each point~$p$ from~$\bC \cup
\{\infty\}$ falls in one of two categories:
\begin{itemize}
\item it is called an \emph{ordinary point of~$L$\/} if the
  coefficient of the highest derivative in the equation does not
  vanish at~$p$, in which case a solution~$f$ is fully determined by
  the initial conditions $f(p)$, \dots, $f^{(r-1)}(p)$ (Cauchy
  theorem);
\item it is called a \emph{singular point of~$L$\/} otherwise.
\end{itemize}

Among singular points, the more tractable situation
% both from the viewpoint of analysis and from the viewpoint of symbolic manipulations
is that of a \emph{regular singular\/} point (of~$L$).  At such a
point, the growth of solutions is bounded (in any small sector) by an
algebraic function.  In the neighbourhood of a regular singular
point~$p$, any solution can be written as a linear combination of
products of a (possibly multivalued) term of the
form~$(x-p)^e$ with~$e\in\bC$, logarithms $\ln(x-p)$,  
(respectively~$(1/x)^e$ and $\ln(1/x)$
if~$p=\infty$) and analytic (univalued) functions.  The
constants~$e$ are called the \emph{exponents\/} of~$L$ at~$x=p$.  They
are the zeroes of the so-called \emph{indicial equation\/} of~$L$
at~$x=p$, an algebraic equation that is defined for any $L$ and~$p$,
and can be obtained algorithmically.

Another simple situation occurs at a singular point~$p$ of~$L$ when
there exist a function~$u$ (possibly singular at~$p$) and functions
$y_1$, \dots, $y_r$ that are analytic in a neighbourhood of~$p$, such
that $uy_1$, \dots, $uy_r$ constitute a basis of the solution space
of~$L$.  Then, we say that $p$~is a \emph{removable singularity\/}
of~$L$.  Otherwise, we say that it is a \emph{non-removable singularity}.

Assume from now on that $L$~is irreducible and of order 2.  Then, the
indicial equation at~$x=p$ has degree at most~2; in other words, there
are at most~2 exponents at~$x=p$ (counting with multiplicity).  We
shall also assume that $L$~is regular singular at~$x=p$, which implies
that the degree of the indicial equation is equal to~2 and that there
are two exponents, $e_1$ and~$e_2$ (counting with multiplicity).  The
\emph{exponent difference\/}~$d$ at $x=p$ is defined up to sign as
$\pm (e_1 - e_2)$.

The numbers $e_1$,~$e_2$, and~$d$ describe a lot of the asymptotic
behaviour of the solutions of~$L$ when $x$~tends to~$p$.  Let us be
more specific about this.  For notational convenience, consider first
the case $p=0$.  Then $e$~is an exponent at $x=0$ if and only if $L$
has a formal solution $u$ for which $u/x^e \in R \setminus xR$, where
$R := \bC[[x]][\ln(x)]$.  If $\ln(x)$ actually occurs in~$u/x^e$ then
the point~$p$ is called \emph{logarithmic}.  If $\ln(x)$ does not
occur, then our assumption that $L$~is regular singular can be used to
prove that $u/x^e$~is analytic at~$x=0$, so that $x^e$~describes the
local behaviour of~$u$ at~$x=0$.  Similar statements hold for $p \neq
0$, after replacing $x$ by~$x-p$ if $p$~is finite, or $x$ by~$1/x$
if~$p=\infty$.  The following lemma summarises the link
between the exponent difference of a second-order operator and the
asymptotic behaviour of its solutions. 
\begin{lem}
Let $L$~be a linear difference operator of order~2 with exponent
difference~$d$ at some point $p \in \bC \cup \{\infty\}$.  Assume
$p$~to be either an ordinary or a regular singular point of~$L$.
Then:
\begin{enumerate}
\item If $p$ is ordinary, then $d=1$.
\item If $p$ is a removable singularity, then $d \in \bZ
  \setminus \{0\}$.
\item If $p$ is logarithmic, then $d \in \bZ$.
\item If $d=0$, then $p$~is logarithmic.
\item If $d \in \bZ \setminus \{0\}$, then $p$ is either logarithmic
  or a removable singularity.
\end{enumerate}
\end{lem}

For proofs, see \cite[Theorem~1]{debeerst}, or see \cite{RubenMasters} for more details.

\subsubsection{Transporting the singularities of the Gauss
  hypergeometric equation}

Let $L^{a,b}_c$ denote the differential operator of the Gauss
hypergeometric equation,
\[ x(1-x)\frac {d^2w}{dx^2} + \bigl(c-(a+b+1)x \bigr) \frac{dw}{dx} - abw = 0. \]
Of its two linearly independent classical solutions,
we shall only need one: $\pFqnoargs21(a,b;c;x)$.
It has 3~singular points: 0, 1, and~$\infty$.
The respective exponent differences are $(e_0, e_1, e_{\infty}) =
(\pm(1-c), \pm(c-a-b), \pm(a-b))$.

Let $f$ be a rational function, and denote by $L^{a,b}_{c;f}$ the
operator obtained from $L^{a,b}_c$ by the change of variables~$x=f$,
so that it has $\pFqnoargs21(a,b;c;f)$ as a solution.

The following lemma explains how the exponent differences of the Gauss
differential equation get transported by the composition under~$f$. The proof
is the same as in \cite[Theorem~2(i)]{debeerst} (more details are given in
\cite{RubenMasters}).

\begin{lem}
Let $d$ be the exponent difference of $L^{a,b}_{c;f}$ at $x=q$. Then:
\begin{enumerate}
\item If $q$ is a root of $f(x)$ with multiplicity $m$, then $d = m \cdot e_0$. 
\item If $q$ is a root of $f(x)-1$ with multiplicity $m$, then $d = m \cdot e_1$. 
\item If $q$ is a pole of $f(x)$ of order $m$, then $d = m \cdot e_{\infty}$.
\end{enumerate}

\end{lem}

Recall that $L$ is irreducible and of order~2.
Let $y = \pFqnoargs21(a,b;c;f)$ where $f$~is some rational function.
By solving~$L$ in terms of~$y$, we mean finding three rational
functions $r$, $r_0$, and~$r_1$ such that
\begin{equation}\label{form}
\exp\left(\int r\right) \cdot (r_0 y + r_1 y')
\end{equation}
is a solution of $L$ (see \cite[Definition~2,
  parts~\emph{(ii),(iii)\/}]{debeerst}).  A necessary condition for such
a solution to exist is that $L$~has the same set of non-removable
singularities as $L^{a,b}_{c;f}$, and moreover, that the exponent
differences match modulo~$\bZ$ (see \cite[Lemma~4]{debeerst}).  The
expression $\exp(\int r)$ is usually a radical function, i.e., a
function whose $r$th power is rational for some integer~$r$.

Consider the operator $P_2$ in~\eqref{eq:P2}.  Its non-removable
singularities are $0$, $1$, $1/64$, $2/3$, and~$\infty$, and all these
singular points are logarithmic; on the other hand, the root~$x=-1/6$ of
the leading coefficient of~$P_2$ is a removable singularity.

Suppose that we choose $a,b,c$ so that $(e_0,e_1,e_{\infty}) =
(0, 1/3, 0)$ (this choice will be explained in Section~\ref{subsubsimp}),
and want to solve $P_2$ in terms of $y := \pFqnoargs21(a,b;c;f)$.  
By the discussion above, if
$p$ is a logarithmic point for~$P_2$ then it must be a logarithmic
point for~$L^{a,b}_{c;f}$ and vice versa.
Since $x=0$ and $x=\infty$ are the logarithmic points of $L^{a,b}_c$ one sees that
the union of the set of roots and the set of poles of $f$
must be exactly $\{0,1,1/64,2/3,\infty\}$.
For each possible degree of~$f$, this leaves finitely many candidate
integer values for
$n_0$, $n_1$, $n_{1/64}$, and~$n_{2/3}$, such that $f$~is of the form
$c (x-0)^{n_0} (x-1)^{n_1} (x-1/64)^{n_{1/64}} (x-2/3)^{n_{2/3}}$
for an unknown constant~$c$.
For an exhaustive search, we therefore try successive potential
degrees $2,3,4,\ldots$ for~$f$.

The exponent difference of $L^{a,b}_{c;f}$ at $x=q$ is $m \cdot e_1$ if $q$ is
a root of $f-1$ with multiplicity $m$.  So $m \cdot e_1$ must match
(modulo~$\bZ$) an exponent difference of $P_2$.
Since $P_2$ has only integer exponent differences, and $e_1 = 1/3$, it
follows that 3~divides~$m$ and so the numerator of $f-1$ must be a cube.
Thus, we try various $n_0,n_1,\ldots \in \bZ$, and for each, we check if there exists $c \in \bC$
for which the numerator of $c (x-0)^{n_0} (x-1)^{n_1} (x-1/64)^{n_{1/64}}(x-2/3)^{n_{2/3}} - 1$ is a cube.
A computer implementation quickly finds 
\[ n_0 = 1,\quad n_1 = -2,\quad n_{2/3} = 1,\quad n_{1/64} = -1,\quad c=-81/64, \]
and we get
\[ f = \frac{-81x(x-2/3)}{64 (x-1)^2(x-1/64)}. \]
Next, we compute $a,b,c$ for which the exponent differences $e_0,e_1,e_{\infty}$ are $0, 1/3, 0$
and find $L^{a,b}_c$.  Applying a change of variables produces $L^{a,b}_{c;f}$.  Then a call to the program \verb+equiv+
\cite{URLequiv}
provides a map of the form $y \mapsto \exp(\int r)(r_0 y + r_1 y')$
from the solutions of $L^{a,b}_{c;f}$ to the solutions of $P_2$.  This way we find a solution of $P_2$
in the form~\eqref{form}.

\subsubsection{Simplifying the obtained solution}
\label{subsubsimp}

The question now is how to make this solution more compact.  First of all, the numerator of $f-1$ is a cube.
Interchanging $e_1$ and~$e_{\infty}$ makes the denominator of~$f$ a
cube.
That makes the expression for $f$
slightly smaller. Now, composition with $x/(x-1)$ interchanges $1,\infty$ while keeping $0$ fixed, so we replace $f$ by $f/(f-1)$
and obtain a new $f$,
\[ f = \frac{27x(2-3x)}{(1-4x)^3}, \]
which corresponds to the new $(e_0,e_1,e_{\infty}) = (0,0,1/3)$.
Next, we try to change $e_0,e_1,e_{\infty}$ by some integers in such a way that the exponent differences
of $L^{a,b}_{c;f}$ not only match those of~$P_2$ modulo~$\bZ$, but are also as close as possible, because this
tends to reduce the size of $r_0, r_1$.
Surprisingly, one can even match the exponent differences exactly! That allows $r_1$ to become 0. We get an
exact match for $(e_0,e_1,e_{\infty}) = (1,1,1/3)$,
corresponding to $a=1/3, b=2/3, c=2$,
and end up with the following solution
\[ \frac{6}{(1-4x)(1-64x)} \cdot \pFq21{1/3}{2/3}{2} { \frac{27
x(2-3x)}{(1-4x)^3} }. \]
This expression is not unique because we can interchange $(e_0,e_1,e_{\infty})$, updating $f$
accordingly. We can also change $e_0,e_1,e_{\infty}$
by integers (as long as those integers add up to an even number) but doing that would make $r_1$ non-zero again.
Moreover, there are other exponent differences that would have solved $P_2$ as well.

The set of $(e_0,e_1,e_{\infty})$'s (modulo $\mathbb{Z}$) we should try is given in
\cite[Table~(I), Section 4]{takeuchi}.
To make the expression size as small as possible, the $(e_0,e_1,e_{\infty})$ that
leads to the lowest degree for~$f$ should be tried first.
The table shows how the degrees
are related to each other; to minimise the degree, we should check $(0,0,0)$ first, and
$(0,1/2,1/3)$ last.  The first case $(0,0,0)$ did not lead to a solution.
The second case we tried was $(0,0,1/3)$, and since that led to a
solution, we did not need to try more cases.
The table in \cite{takeuchi} shows some of the other cases must also lead to a solution; one sees in the table
that $(0,0,1/3)$ (and hence $P_2$) is solvable in
terms of $(0,1/2,1/6)$ and $(0,1/2,1/3)$.
We can also read from the table that if we had used those exponent
differences for our $\pFqnoargs21$, then the degree of~$f$ would have
been 2, respectively~4, times higher.

\bigskip
A \textsf{Maple} session to perform all operations of this section is
given online at
\url{http://algo.inria.fr/chyzak/Rooks/Compute2F1.mpl}.

\section{Further comments}\label{sec:further}

\subsection{Asymptotics}

It was pointed out in~\cite{ErFeTr10} that general results from~\cite{RaWi07} on
the asymptotics of diagonal coefficients of multivariate generating functions
imply
\begin{equation} \label{eq:asympt}
a_n
\sim \frac{9 \sqrt{3}}{40 \pi} \cdot \frac{64^n}{n}.
\end{equation}
Here we show that the same result can be derived in a simpler way (using \emph{univariate}, instead of \emph{multivariate}, singularity analysis), based on the explicit expression~\eqref{eq:closedform}.

Indeed, basic singularity analysis~\cite{FlOd90,FlSe09} shows that, since the dominant singularity of $\dx(\Diag(f))$ is $x=1/64$, with residue
\[
r = \frac{6}{(1-\frac{4}{64})} \cdot \pFq21{1/3}{2/3}{2} { \frac{\frac{27}{64}(2-\frac{3}{64})}{(1-\frac{4}{64})^3}}
=
\frac{32}{5} \cdot 
{\pFq21{1/3}{2/3}{2}{1}}, 
\]
the asymptotic behaviour of the coefficient $h_n$ of $\dx(\Diag(f))$ is $h_n \sim r \cdot 64^n$,
and thus
$\displaystyle{a_n \sim \frac{r}{64} \cdot \frac{64^n}{n}}$.

The value of $r$ is found using Gauss's formula
\[
\pFq21{a}{b}{c}{1} =
\frac{\Gamma(c)\Gamma(c-a-b)}{\Gamma(c-a)\Gamma(c-b)}
\]
and the multiplication formula
\[
\Gamma(z) \cdot \Gamma\left(z + \frac{1}{m}\right)
\cdots \Gamma\left(z + \frac{m-1}{m}\right) = (2 \pi)^{(m-1)/2} \; m^{\frac12 - mz} \; \Gamma(mz)
\]
with $m=3$ and $z=1$.
This gives $r = {72}{\sqrt{3}}/(5 \pi)$,
and proves~\eqref{eq:asympt}.

\subsection{Non-algebraicity and simpler formulas}

One may wonder whether the function $\Diag(f)$ is algebraic. This is not the
case, and can be seen in several ways:
\emph{(i)\/}~as a consequence of the asymptotic
behaviour of the coefficients $a_n$ (generating series of sequences
asymptotically equivalent to $n^\alpha\rho^n$ with $\alpha \in \bZ_{-}$ are
necessarily transcendental~\cite[Thm~D]{Flajolet87});
\emph{(ii)\/}~by doing a local
analysis of $P$ near its singularities (the presence of logarithmic terms in a
local expansion excludes algebraicity); \emph{(iii)\/}~by applying Schwarz's
classification~\cite{Schwarz1873} of algebraic~$_2F_1$'s.

It is also legitimate to seek for an even simpler closed form, not
involving any integration sign. Such a reduction is not possible:
if it were, the algorithm given in~\cite{integrateSols} should have found it.
The sequence $(a_n)$ satisfies the third-order
recurrence~\eqref{eq:shortrec}, but no second-order recurrence with
polynomial coefficients, for otherwise \eqref{eq:shortrec} or a
closely related adjoint recurrence would need to possess a
hypergeometric solution, which is easily sorted out by Petkov\v{s}ek's
algorithm~\cite{Petkovsek-1992-HSL}.

\subsection{An alternative hypergeometric expression}
Using a completely different method, Frits Beukers~\cite{Beukers10}
independently found that the diagonal 3D rook path series $G(x)$ satisfies
\begin{equation}\label{Beukers}
	G'(x) = \frac{1-x}{2(1+6x)}\biggl( (1-4x) H'(x) - 4 H(x) \biggr), 
\end{equation}	
where \[H(x) = \frac{1}{g_2^{1/4}} \cdot \pFq21{1/12}{5/12}{1}{\frac{1}{J(x)}},\]
with
\begin{multline*}
g_2 = (1-4x) (1-60x+120x^2-64x^3), \\
 J(x) = \frac{g_2^3}{1728(1-x)^2 x^3 (2-3x)^3 (1-64x)}.
\end{multline*}
The function $J(x)$ is obtained, by a method similar to the one in~\cite{Beukers82}, as the $J$-invariant of a family of elliptic curves naturally associated to the integral interpretation of the diagonal problem.

\medskip
Of course, once obtained, it is an easy task to check that the new
hypergeometric expression coincides with ours: the difference of the
two hypergeometric expressions satisfies a simple linear differential
equation derived from the defining linear differential equation for
the hypergeometric~$\pFqnoargs21$, and it is easily expanded as a
Taylor series, showing that enough first terms are zero, so that the
difference is zero. This calculation is available as Maple code at
\url{http://algo.inria.fr/chyzak/Rooks/CompareFormulas.mpl}.

\medskip
Interestingly, one can algebraically reduce one expression to the
other by combining only two classical hypergeometric identities:
\begin{itemize}
	\item[(a)] Gauss' contiguity relation
\begin{equation}\label{Gauss}
 \frac{9(1-x)}{2} \cdot \pFq21{1/3}{2/3}{1}{x}' =  \pFq21{1/3}{2/3}{2}{x},
\end{equation}
obtained by evaluating at 
$a=\frac13, b=\frac23$, and $c=1$
the more general contiguous transformation
\begin{multline*}
c(1-x) \cdot \frac{d}{dx}\, \pFq21{a}{b}{c}{x} = \\
(c-a)(c-b) \cdot \pFq21{a}{b}{c+1}{x} + c(a+b-c)\cdot \pFq21{a}{b}{c}{x},
\end{multline*}
\item[(b)] Goursat's quartic relation
\begin{equation}\label{Goursat}
 \pFq21{1/3}{2/3}{1}{x} = \pFq21{1/12}{5/12}{1}{\frac{64x^3(1-x)}{(9-8x)^3}}
 \cdot  \biggl(1 - \frac{8x}{9} \biggr)^{- 1/4}, 
\end{equation}
obtained by taking $\alpha=\frac{1}{12}$ in the identity~\cite[Eq.~(126)]{Goursat1881} (see also~\cite[Eq.~(25)]{Vidunas09}):
\[\pFq21{4\alpha}{4\alpha+\frac13}{6\alpha+\frac12}{x} = \pFq21{\alpha}{\alpha+\frac13}{2\alpha+\frac56}{\frac{64x^3(1-x)}{(9-8x)^3}}
 \cdot \biggl(1 - \frac{8x}{9} \biggr)^{3\alpha}. \]
\end{itemize}

\medskip
The detailed reduction is as follows:
Equations~\eqref{Gauss} and~\eqref{Goursat} show that 
\[
\pFq21{1/3}{2/3}{2}{x} = \frac{9(1-x)}{2} \cdot \biggl( \pFq21{1/12}{5/12}{1}{\psi(x)}
\cdot \chi(x) \biggr)',
\]
where $\psi(x) = {64x^3(1-x)}/{(9-8x)^3}$ and $\chi (x)  = \bigl(1 - \frac{8x}{9} \bigr)^{- 1/4}$. By substitution, the latter equality implies 
\begin{multline*}
\pFq21{1/3}{2/3}{2}{\varphi(x)} = \\
\frac92 \cdot \frac{1-\varphi(x)}{\varphi'(x)} \cdot \biggl( \pFq21{1/12}{5/12}{1}{\psi (\varphi(x))}
\cdot \chi (\varphi(x)) \biggr)',
\end{multline*}
with $\varphi(x) = {27x(2-3x)}/{(1-4x)^3}$. The right-hand side term rewrites
\[
\frac{(1-64x)(1-x)(1-4x)}{12 (1+6x)} \cdot \biggl( \pFq21{1/12}{5/12}{1}{\frac{1}{J(x)}}
\cdot \left(\frac{g_2}{(1-4x)^4}\right)^{-1/4} \biggr)',
\]
and this, in combination with equation~\eqref{eq:closedform}, finally implies equation~\eqref{Beukers}.

\subsection{The case of 3D queens}

A closely related problem is to find the generating function for the diagonal
sequence for 3D queen paths. This is Sequence A143734 in Sloane's
encyclopedia \url{http://oeis.org/A143734}, whose
first terms read
\[1, 13, 638, 41476, 3015296, 232878412, 18691183682, 1540840801552, \ldots \]

Computationally, it turns out that this problem is more difficult than the one
on 3D rooks.  By Lemma~\ref{basiclemma}, it boils down to finding the
complete diagonal $Q(x)$ of the rational function 
\[ 
\left(1 - \frac{s}{1-s} - \frac{t}{1-t} - \frac{x}{1-x} - \frac{st}{1-st} -\frac{tx}{1-tx} - \frac{xs}{1-xs} -\frac{stx}{1-stx}  \right)^{-1}.
\]

We empirically found (using guessing routines based on Hermite--Padé
approximants) that $Q(x)$ satisfies a differential equation $L_Q(x,\dx)$ of
order~6 and polynomial coefficients (in $x$) of degree at most~71. The
corresponding sequence satisfies a recurrence of order~14 with polynomial
coefficients (in~$n$) of degree at most~52. Moreover, it is very likely that
these are the \emph{minimal-order\/} equations. They are available
online at \url{http://algo.inria.fr/chyzak/Rooks/QueensRecEq.mpl} for the
recurrence and \url{http://algo.inria.fr/chyzak/Rooks/QueensDiffEq.mpl} for
the differential equation.

The conjectured equations already have a large size, so that proving
their correctness by the creative-telescoping approach becomes a
challenge; we have not yet succeeded in proving this in a fully
algorithmic fashion.

However, we are convinced that our guessed equations are correct: they pass
all the various check tests described in~\cite{BoKa09}. In addition, the
degree-71 leading coefficient of the differential equation factors as
\begin{multline*}
x^2 \cdot (5x-4) \cdot (x+1) \cdot (x-1)^6 \cdot (36x^2-40x+9) \cdot {} \\
(512x^4-661x^3+84x^2+95x-1) \cdot {} \\
(14063 x^6-15940 x^5-5918 x^4+12063 x^3-4118 x^2+575 x-40) \cdot P(x)
\end{multline*}
where $P(x) \in \bQ[x]$ is irreducible of degree~49.

Note the known fact~\cite{RaWi07, ErFeTr10} that the $n$th coefficient of
$Q(x)$
is asymptotically equal to $\kappa c^{-3n}/n$, for some $\kappa \in \bR$
and $c \approx 0.2185$.
More specifically, $c$~is the smallest positive real root of the
equation
\[ 1 - \binom31 \frac{x}{x-1} - \binom32 \frac{x^2}{1-x^2} - \binom33 \frac{x^3}{1-x^3} = 0 , \]
which is consistent with  the presence of the factor $512x^4-661x^3+84x^2+95x-1$
having $c^3 \approx 0.0104$ as a root (in fact, its root that is closest to zero).

The sixth-order differential operator~$L_Q(x,\dx)$ can be factored as $L_Q=L_4 \cdot R_1^{(a)} \cdot R_1^{(b)}$, where $L_4$ has order 4, and both $R_1^{(a)}$ and $R_1^{(b)}$ have order 1.
By analogy with our result on rook paths, a natural question is whether~$L_4$ is solvable
in terms of indefinite integrals, algebraic functions, and $\pFqnoargs21$'s expressions.
If it were, then it would be solvable in terms of solutions of second-order equations,
and by \cite{Singer85} at least one of the following should then be true (see \cite[Sections~1.1 and~1.2]{ReduceOrder} for definitions and details):
\begin{enumerate}
\item $L_4$ is reducible,
\item $L_4$ is gauge equivalent
(under Maple's \textsf{DEtools[Homomorphisms]}) 
to the third symmetric power of a second-order operator,
\item $L_4$ is gauge equivalent to the symmetric product of two second-order operators.
\end{enumerate}
Regarding option~(1), the operator does not factor (\textsf{DFactor} in Maple).  Regarding options~(2) and~(3),
\cite[Section~3]{ReduceOrder} shows how to find order reductions of such type (if they exist).
We found that no such reduction exists. Option~(2) can be ruled out quickly by local data
(if a third symmetric power has a logarithmic solution at $x=p$, then the cube of a logarithm should appear as well).
For option~(3), we computed the eigenring of the second exterior power of $L_4$.
To conclude, the fourth-order factor~$L_4$ is not solvable (in the
sense of Singer's definition~\cite{Singer85}) in terms of solutions of
second-order equations;
in particular, this means that, in this definition, there is no expression in
terms of $\pFqnoargs21$ functions for the generating function $Q(x)$ of
diagonal 3D queen paths.

It appears that $L_4$ has no $\pFqnoargs43$ type solution either, but
we have not obtained a proof of this fact yet.

\subsection{Faster but heuristic approach to getting the certificates}

As already mentioned in the introduction, we have chosen in
Section~\ref{sec:solution} to confine ourselves to calculations that follow an
algorithmic approach, that is, that would provably (theoretically)
succeed on any other rational input.  Beside this, heuristics are
available to cleverly guess a solution to the key equation, including
certificates, and often lead to faster computations when they succeed,
although they can fail in theory and sometimes do in practice.  One of
the most advanced strategies in this spirit is due to
Koutschan~\cite{Koutschan-2010-FAC}.  Note that by rational-function
normalisation, verifying the key equation a~posteriori \emph{proves\/}
the obtained results.

We have implemented such a strategy in the rational case, and we could
get a solution~$(P, S', T')$, where $P$~is the same operator as in
Section~\ref{sec:the-result}, but for other certificates $S'$
and~$T'$.  In effect, this heuristically solves a partial differential
equation for (some of) its rational solutions.  The calculation is
4~times faster than in the algorithmic way.  Interestingly enough, the
denominators of $S'$ and~$T'$ only involve~$q_1$, and not its
discriminant~$\disc_t(q_1)$.

\subsection*{Acknowledgements} We wish to thank the referee for his quick and
serious job and for his constructive comments  on the
manuscript.

\bibliographystyle{plain}
%\bibliography{BoChHoPe11}
%%%%% Beginning of inserted BoChHoPe11.bbl

%%%%% End of inserted BoChHoPe11.bbl

\appendix

\section{Details of  calculations of the iterated creative telescoping for the 3D rook generating series}
\label{sec:ct-appendix}

In this appendix, the calculation sketched in~\S\ref{sec:sketch} is
explained in more detail, and we provide explicit values of
intermediate computations.
We start by introducing the notions needed for our description.

\subsubsection*{Differential operators}

Whereas the sketch of the algorithm was given with references to the
function~$F$ to be integrated and its derivatives, here we prefer
presenting the calculations as they are done in actual implementations:
a primal role is played by \emph{differential operators\/} that
represent linear combinations of derivatives of a given function.

As certain operators map the function under consideration to zero,
operators have to be viewed up to addition of such annihilating
operators.
This explains the use of congruences in the calculations below,
denoted $L_1 \equiv L_2 \mod I$, when $L_1(F) = L_2(F)$ and $I$~is the
(left) ideal of all operators that annihilate~$F$.

\subsubsection*{Representing a D-finite function}

Recall from~\cite{Stanley-1980-DFP,Lipshitz-1989-DFP} the definition
that a D-finite function of variables $u$ and~$v$ is a function~$F$
for which there exists a finite maximal set of
derivatives~$\du^a\dv^b(F)$ that are linearly independent
over~$\bQ(u,v)$.
The natural question is how a function is algorithmically recognised
and represented as D-finite.

The best algorithmic answer would be in terms of non-commutative
Gröbner bases (and ideals of Hilbert dimension~0)
\cite{Chyzak:2009:NHS}, but here, to keep the sophistication to a
minimum, we shall use the alternate definition that a function is
D-finite if there are non-zero operators $A(u,v,\du)$ and~$B(u,v,\dv)$
that annihilate it, one for each derivative.
Following~\cite{Chyzak:2009:NHS}, the pair~$(A,B)$ of annihilating
operators will be called a \emph{rectangular system}.

\subsubsection*{Rational solutions of parametrised systems of linear
  differential equations}

Chyzak's algorithm relies on an algorithm that, given matrices $A\in
\cM_n(\bK(t))$ and $B\in \cM_{n\times d}(\bK(t))$ for some field~$\bK$
that depends on the application, solves
\begin{equation}\label{eq:param-lin-sys}
\frac{\partial y}{\partial t}+Ay=Be
\end{equation}
for $y\in \bK(t)^n$ and $e\in\bK^d$.
In our uses, the field~$\bK$ is a rational function field in the 
other variables of the problem.
We do not describe this algorithm here, and refer the reader for
instance to~\cite{Abramov:1991:FAS} (and the description of its use
in~\cite{Chyzak:2000:EZF}).

\bigskip

For convenience, let us recall the notation of~\S\ref{sec:the-result}.
Let $q$ be the denominator of $F$, then $q=stq_1$ with
$q_1=-st+2s^2t+2t^2+2xs-3st^2-3xt-3xs^2+4xst$ an irreducible
polynomial of $\bQ[x,s,t]$.
Let $\disc_t(q_1)=(x-s)(16xs^2-4s^3-24xs+4s^2+9x-s)$ be the
discriminant of $q_1$ with respect to $t$.

\subsection{Stage~A: First iteration of Chyzak's algorithm}

Stage~A in~\S\ref{sec:sketch} looks for operators
$P^{(\alpha)}\in\bQ(x,s)\langle\dx,\ds\rangle$
(according to the ansatz~\eqref{eq:P-Ansatz-2})
and rational
functions $\phi^{(\alpha)}\in\bQ(x,s,t)$ such that
$P^{(\alpha)}(F)=\frac{\partial}{\partial t}(\phi^{(\alpha)}F)$
(cf.~\eqref{eq:identities-after-first-iteration}).
A property of the left ideal $\langle P^{(\alpha)}\rangle$
thus obtained in $\bQ(x,s)\langle\dx,\ds\rangle$ is that it
contains a rectangular system.

\subsubsection*{Ansatz with $r=1$.}

We first look for operators $P^{(\alpha)}$ of total degree~$r=1$.
Let $\eta_{0,0},\eta_{1,0},\eta_{0,1}\in\bQ(x,s)$ be
unknowns such that $P=\eta_{0,0}+\eta_{1,0}\dx+\eta_{0,1}\ds$,
and let $\phi\in\bQ(x,s,t)$ be another unknown.
The equation $P(F)=\frac{\partial}{\partial t}(\phi F)$ rewrites
\begin{equation*}
\eta_{0,0}F+\eta_{0,1}\frac{\partial F}{\partial s}+\eta_{1,0}\frac{\partial F}{\partial x}=
\frac{\partial\phi}{\partial t}F+\phi\frac{\partial F}{\partial t} ,
\end{equation*}
which is an equation in $\phi$ and
$\eta_{0,0},\eta_{0,1},\eta_{1,0}$, of the
form~\eqref{eq:param-lin-sys} with~$\bK=\bQ(x,s)$.
All solutions are proportional to the one given by
\begin{equation}\label{eq:eta-and-phi-for-P_1}
\left\{\begin{aligned}
\eta_{0,0}&=2(s-1)(3s^2-6s+2),\\
\eta_{1,0}&=6xs^3-2s^3-10xs^2+s^2-4x^2s+10xs+3x^2-4x,\\
\eta_{0,1}&=2s(3s-2)(s-1)^2,\\
\phi&=\frac{-t(s-1)(-6xs^2+6s^2t-s^2+11xs-9st-4x-xt+4t)}{t-x}.
\end{aligned}\right.
\end{equation}
Set $P_1$ to the operator~$P$ obtained for these specific values.
The ideal generated by it obviously does not contain any rectangular
system in $\bQ(x,s)\langle\dx,\ds\rangle$.

\subsubsection*{Ansatz with $r=2$.}

Since all the monomials~$\ds^i$ can be reduced by~$P_1$, we set~$r=2$
and look for
$P=\eta_0+\eta_1\dx+\eta_2\dx^2$, with
$\eta_0,\eta_1,\eta_2\in\bQ(x,s)$, and
$\phi\in\bQ(x,s,t)$ such that
$P(F)=\frac{\partial}{\partial t}(\phi F)$.
(This ansatz is not strictly of the form~\eqref{eq:P-Ansatz-2} that
would involve~$\ds$; this will require additional manipulations when
obtaining a rectangular system.)
The equation on~$F$ becomes
\begin{equation*}
\eta_0F+\eta_1\frac{\partial F}{\partial x}+\eta_2\frac{\partial^2 F}{\partial x^2} =
\frac{\partial\phi}{\partial t}F+\phi\frac{\partial F}{\partial t} ,
\end{equation*}
which is a parametrised equation in $\phi$ and
$\eta_0,\eta_1,\eta_2$, with $\bK=\bQ(x,s)$.
All solutions are proportional to the one given by:
\begin{equation}\label{eq:eta-and-phi-for-P_2}
\left\{\begin{aligned}
\eta_0&= 0 , \\
\eta_1&= -2(-19s^2-9x+13s^3+7s-16xs^2+24xs) , \\
\eta_2&= \disc_t(q_1) , \\
\phi&= \frac{-t(3s-2)(s-1)(2s^2-4st-s+3t)(s-t)^2}{(t-x)q_1} ,
\end{aligned}\right.
\end{equation}
and we call~$P_2$ the operator~$P$ obtained for these specific values.
The ideal $\langle P_1,P_2\rangle$ now contains a rectangular system:
eliminating~$\dx$ from $P_1$ and~$P_2$ results in an operator
$P'_2=\eta'_0+\eta'_1\ds+\eta'_2\ds^2$, with coefficients of respective
degrees $(4,9)$, $(4,10)$, and $(4,11)$ in~$(x,s)$.
This ends the calculation.

The Maple code to compute~$P'_2$ can be obtained at
\url{http://algo.inria.fr/chyzak/Rooks/RectangularSystem.mpl}.

\subsection{Stage~B: Second iteration of Chyzak's algorithm}

To perform Stage~B in~\S\ref{sec:sketch},
we now work modulo the ideal $I=\langle P_1,P_2\rangle$ of
$\bQ(x,s)\langle\dx,\ds\rangle$.
The vector space $V=\bQ(x,s)\langle\dx,\ds\rangle/I$ is 2-dimensional
over~$\bQ(x,s)$, with basis $(1+I,\dx+I)$.
In other words, we consider a generic solution~$\hat F$ of $P_1$
and~$P_2$, and represent it by~$1+I$.

To simplify the notation in what follows, we normalise the operators
into the form
$P_1 / (2s(3s-2)(s-1)^2) = \ds-p_1\dx-p_0$ and
$P_2 / \disc_t(q_1) = \dx^2-q\dx$, for coefficients
$p_0,p_1,q\in\bQ(x,s)$.

The algorithm now starts from ansatz~\eqref{eq:P-Ansatz-1} for~$P$, but
with the form~\eqref{eq:S-as-finite-sum} for~$S$, in order to find a
relation of the form~\eqref{eq:identities-after-second-iteration}.
For increasing values of $d$, it looks for an operator
$P=\sum_{i=0}^d\eta_i\dx^i$, such that $P\equiv\ds Q\mod I$, with
$\eta_i\in\bQ(x)$, $\phi_0,\phi_1\in\bQ(x,s)$, and
$Q=\phi_0+\phi_1\dx$.

There is no solution for $d<3$, and we give details for~$d=3$ only.
First, the algorithm simplifies~\eqref{eq:bivariate-rat-ct}.
Modulo~$I$, we have $\dx^2\equiv q\dx$, and therefore
\[\dx^3 \equiv \frac{\partial q}{\partial x}\dx+q\dx^2 \equiv
\left(\frac{\partial q}{\partial x}+q^2\right)\dx.\]
Additionally, $\ds\equiv p_0+p_1\dx$, so
\[
\dx\ds
  \equiv \frac{\partial p_0}{\partial x}+\left(p_0+\frac{\partial p_1}{\partial x}\right)\dx+p_1\dx^2
  \equiv \frac{\partial p_0}{\partial x}+\left(p_0+\frac{\partial p_1}{\partial x}+p_1q\right)\dx.
\]
As to the right-hand side:
\begin{multline*}
\ds Q=\frac{\partial\phi_0}{\partial s}+\phi_0\ds+\frac{\partial\phi_1}{\partial s}\dx+\phi_1\ds\dx\\
\equiv \frac{\partial\phi_0}{\partial s}+\left(p_0+p_1\dx\right)\phi_0+\frac{\partial\phi_1}{\partial s}\dx
  +\phi_1\left(\frac{\partial p_0}{\partial x}+\left(p_0+\frac{\partial p_1}{\partial x}+p_1q\right)\dx\right).
\end{multline*}
Since $P \equiv \eta_0+\left(\eta_1+q\eta_2+\left(\frac{\partial
q}{\partial x}+q^2\right)\eta_3\right)\dx$, the congruence
$P\equiv\ds Q\mod I$ expressed in the basis $(1+I,\dx+I)$ is
equivalent to the differential system

\begin{equation}\label{eq:param-sys}
\left\{\begin{aligned}
\frac{\partial\phi_0}{\partial s}&=-p_0\phi_0-\frac{\partial p_0}{\partial x}\phi_1+\eta_0,\\
\frac{\partial\phi_1}{\partial s}&=
-p_1\phi_0
-\left(p_0+\frac{\partial p_1}{\partial x}+p_1q\right)\phi_1
+\eta_1+q\eta_2+\left(\frac{\partial q}{\partial x}+q^2\right)\eta_3.
\end{aligned}\right.
\end{equation}
Solving this parametrised system of linear differential equations
in~$\bQ(x)$ results in a 1-dimensional vector space, with basis given
by the solution
\begin{equation}\label{eq:eta-and-phi-for-final-P}
\left\{\begin{aligned}
\eta_0&= 0,\\
\eta_1&= 4(576x^3-801x^2-108x+74),\\
\eta_2&= 4608x^4+813x^2-6372x^3+514x-4,\\
\eta_3&= x(x-1)(64x-1)(3x-2)(6x+1),\\
\phi_0&= \frac{(53+108x)(3s-2)s}{s-1},\\
\phi_1&= \frac{\gamma}{2(s-1)^2\disc_t(q_1)},\\
\end{aligned}\right.
\end{equation}
where $\gamma$ is an irreducible polynomial in $\bQ[x,s]$ of degree $(5,7)$
in $(x,s)$.

The Maple code to compute~$\gamma$ can be obtained at
\url{http://algo.inria.fr/chyzak/Rooks/SecondIteration.mpl}.

\subsection{Stage~C: Reconstruction}

We finally derive an explicit identity of the
form~\eqref{eq:identity-after-two-iterations} in Stage~C
in~\S\ref{sec:sketch} and obtain the corresponding values for $S$
and~$T$ given by~\eqref{eq:recombining-S-and-T}, thus leading to the
solution~$(P,S,T)$ of~\eqref{eq:trivariate-rat-ct}.

The previous calculation of $P_1$ and~$P_2$ can be reinterpreted as
follows:
Introducing the left ideal~$J$ that annihilates~$F$, that is, the
ideal~$J$ generated by
\[ F\dx - \frac{dF}{dx} ,
\quad F\ds - \frac{dF}{ds} ,
\quad F\dt - \frac{dF}{dt} , \]
we have obtained the congruences
\[ P_1 \equiv \dt\psi_1 \mod J
\quad\text{and}\quad
P_2 \equiv \dt\psi_2 \mod J , \]
for rational functions $\psi_1$ and~$\psi_2$ that are the
functions~$\phi$ in \eqref{eq:eta-and-phi-for-P_1}
and~\eqref{eq:eta-and-phi-for-P_2}, respectively.
Additionally, $P \equiv \ds Q \mod I$ for
$P=\eta_0+\eta_1\dx+\eta_2\dx^2$ and $Q=\phi_0+\phi_1\dx$ described
by~\eqref{eq:eta-and-phi-for-final-P}, and $I=\langle
P_1,P_2\rangle$.

To express $P-\ds Q\in\langle P_1,P_2\rangle$ more explicitly, we look
for $A_1,A_2\in\bQ(x,s)\langle\dx,\ds\rangle$ such that $P-\ds
Q=A_1P_1+A_2P_2$.
This division is made algorithmic by observing that $(P_1,P_2)$ is a
Gröbner basis, and by performing a reduction by this Gröbner basis.
In simple terms, the calculation is as follows:
We first use~$P_1$ to eliminate~$\ds$ from~$P-\ds Q$, then use~$P_2$
to divide the remainder.
Collecting the quotients, we then find:
\[\left\{\begin{aligned}
A_1&=-\frac{108x+53}{2(s-1)^3}+\frac{\gamma_1}{4s(3s-2)(s-1)^4\disc_t(q_1)}\dx,\\
A_2&=\frac{\gamma_2}{4s(3s-2)(s-1)^4\disc_t(q_1)^2}+{}\\
&\qquad\qquad\frac{x(x-1)(64x-1)(3x-2)(6x+1)}{\disc_t(q_1)}\dx,
\end{aligned}\right.\]
with $\gamma_1,\gamma_2$ irreducible polynomials in $\bQ[x,s]$ of
respective degrees $(5,7)$ and~$(7,10)$ in~$(x,s)$.

Since $P_1(F)=\dt(\psi_1F)$ and $P_2(F)=\dt(\psi_2F)$, we get the equality $(P-\ds
Q)(F)=(A_1\dt\psi_1+A_2\dt\psi_2)(F)$. Since $A_1$ and $A_2$ do not
depend on $t$, we get $P(F)=(\ds Q+\dt(A_1\psi_1+A_2\psi_2))(F)$.

Let $S=Q(F)$ and $T=(A_1\psi_1+A_2\psi_2)(F)$. Then $S,T\in\bQ(x,s,t)$
and $P(F)=\frac{\partial S}{\partial s}+\frac{\partial T}{\partial
t}$.
It turns out that $S$ and $T$ are of the very special form
\[S=\frac{(s-t)U}{2stq_1^2\disc_t(q_1)},\qquad T=\frac{(s-t)V}{2s^2q_1^3\disc_t(q_1)^2},\]
with $U,V$ irreducible polynomials in $\bQ[x,s,t]$ of respective
degrees $(5,8,3)$ and $(8,14,5)$ in $(x,s,t)$.

The Maple code for this reconstruction of $S$ and~$T$ can be obtained at
\url{http://algo.inria.fr/chyzak/Rooks/Reconstruction.mpl}.
\end{document}